\documentclass[12pt,preprint]{aastex}

% ...
 
\righthead{Immigrants Welcome}
\lefthead{S.\ F.\ Portegies Zwart et al.}

\begin{document}
\newcommand{\be}{\begin{eqnarray}}
\newcommand{\ee}{\end{eqnarray}}
\newcommand{\etal}{{\it{et al.}}}
\newcommand{\smass}{M_{\odot}}
\newcommand{\br}{{\bf r}}
\newcommand{\bV}{{{\bf v}}}

\newcommand{\LL}{\mbox{${\log \Lambda}$}}
\newcommand{\Ll}{\mbox{${\log \lambda}$}}
\newcommand{\mstar}{m_\star}
\newcommand{\trxJ}{\mbox{${t_{\rm rJ}}$}}
\newcommand{\trxt}{\mbox{${t_{\rm rt}}$}}
\newcommand{\trxh}{\mbox{${t_{\rm rh}}$}}
\newcommand{\tdf}{{\mbox{$t_{\rm df}$}}}
\newcommand{\trh}{{\mbox{$t_{\rm hr}$}}}
\newcommand{\thc}{{\mbox{$t_{\rm hc}$}}}
\newcommand{\trJ}{{\mbox{$t_{\rm Jr}$}}}
\newcommand{\tcc}{{\mbox{$t_{\rm cc}$}}}
\newcommand{\tdis}{\mbox{${t_{\rm dis}}$}}
\newcommand{\aJ}{${a_{\rm J}}$}
\newcommand{\mJ}{{m_{\rm J}}}
\newcommand{\nJ}{{m_{\rm J}}}
\newcommand{\rJ}{{r_{\rm J}}}
\newcommand{\rcore}{{r_{\rm core}}}
\newcommand{\mcore}{\mbox{${m_{\rm core}}$}}
\newcommand{\rvir}{{r_{\rm vir}}}
\newcommand{\aG}{{a_{\rm G}}}
\newcommand{\MG}{{M_{\rm G}}}
\newcommand{\RG}{{R_{\rm G}}}
\newcommand{\lsun}{{L_\odot}}
\newcommand{\mrunaway}{\mbox{${m_{\rm r}}$}}

\renewcommand{\thefootnote}{\alph{footnote}}

\newcommand{\kms}{\mbox{${\rm km~s}^{-1}$}}
\newcommand{\msun}{\mbox{${\rm M}_\odot$}}
\newcommand{\rsun}{\mbox{${\rm R}_\odot$}}
\def\apgt{\ {\raise-.5ex\hbox{$\buildrel>\over\sim$}}\ }
\def\aplt{\ {\raise-.5ex\hbox{$\buildrel<\over\sim$}}\ }
\def\gteq{\ {\raise-.5ex\hbox{$\buildrel>\over-$}}\ }
\def\lteq{\ {\raise-.5ex\hbox{$\buildrel<\over-$}}\ }
\def\gt{\ {\raise-.5ex\hbox{$\buildrel<$}}\ }
\def\lt{\ {\raise-.5ex\hbox{$\buildrel>$}}\ }



\def\steve#1{{\bf[#1 -- Steve]}}
\def\Steve#1{{\bf[#1 -- Steve]}}
\def\simon#1{{\bf[#1 -- Simon]}}
\def\Simon#1{{\bf[#1 -- Simon]}}
\def\holger#1{{\bf[#1 -- Holger]}}
\def\Holger#1{{\bf[#1 -- Holger]}}
\def\Jun#1{{\bf[#1 -- Jun]}}
\def\jun#1{{\bf[#1---Jun]}}
\def\piet#1{{\bf[#1---Piet]}}
\def\Piet#1{{\bf[#1---Piet]}}

%
%
\title{
The ecology of star clusters and intermediate mass
black holes in the Galactic bulge
}
%
%

\author{Simon F.\ Portegies Zwart\altaffilmark{1,2},
 Holger Baumgardt\altaffilmark{3},
 Stephen L.\ W.\ McMillan\altaffilmark{4},
 Junichiro Makino\altaffilmark{5},
 Piet Hut\altaffilmark{6}
\and Toshi Ebisuzaki\altaffilmark{7}
}

\altaffiltext{1}{Astronomical Institute ``Anton Pannekoek,''
	University of Amsterdam, Kruislaan 403, 1098 SH Amsterdam, NL}

\altaffiltext{2}{Department of Computer Science,  University of Amsterdam,
	Kruislaan 403, 1098 SH Amsterdam, NL}

\altaffiltext{3}{Sternwarte, University of Bonn,  Auf dem H\"ugel 71,
	53121 Bonn, Germany}

\altaffiltext{4}{Department of Physics,Drexel University,
	Philadelphia, PA 19104, USA steve@physics.drexel.edu}

\altaffiltext{5}{Department of Astronomy, University of Tokyo, Tokyo 113, Japan}

\altaffiltext{6}{Institute for Advanced Study,
	Princeton, NJ 08540, USA}

\altaffiltext{7}{RIKEN, 2-1 Hirosawa Wako 351-0198, Japan}

\date{Received 2005 August 1; 
      in original form 1687 October 3.6; Accepted xxxx xxx xx.}


\label{firstpage}

%
%

\begin{abstract}
We simulate the inner 100\,pc of the Milky-Way Galaxy to study the
formation and evolution of the population of star clusters and 
intermediate mass black holes. For this study we perform
extensive direct $N$-body simulations of the star clusters which
reside in the bulge, and of the inner few tenth of parsecs of the
super massive black hole in the Galactic center. 
In our $N$-body
simulations the dynamical friction of the star cluster in the tidal
field of the bulge are taken into account via (semi)analytic soluations. 
The $N$-body calculations are used to calibrate a (semi)analytic
model of the formation and evolution of the bulge.

We find that $\sim 10$\% of the clusters born within $\sim100$\,pc of
the Galactic center undergo core collapse during their inward
migration and form intermediate-mass black holes (IMBHs) via runaway
stellar merging.  After the clusters dissolve, these IMBHs continue
their inward drift, carrying a few of the most massive stars with
them.  We predict that region within $\sim10$ parsec of the SMBH is
populated by $\sim 50$~IMBHs of $\sim 1000$~\msun.  Several of these
are expected to be accompanied still by some of the most massive stars
from the star cluster.  We also find that within a few milliparsec of
the SMBH there is a steady population of several IMBHs. This
population drives the merger rate between IMBHs and the SMBH at a rate
of about one per 10Myr, sufficient to build the accumulate majority of
mass of the SMBH.  Mergers of IMBHs with SMBHs throughout the universe
are detectable by LISA, at a rate of about two per week.

\end{abstract}

\section{Introduction}

In recent years the Galactic center has been explored extensively over
most of the electromagnetic spectrum, revealing complex structures and
a multitude of intriguing physical phenomena.  At the center lies a
$\sim3.7\times10^6$ solar mass (\msun) black hole
\citep{1997MNRAS.284..576E,1998ApJ...509..678G,2000Natur.407..349G}.
The presence of a water-rich dust ring at about one parsec from
Sgr\,A* \citep{2003A&A...402L..63S}. further underscores the
complexity of this region, as does the presence within the central
parsec of a few million year old population of very massive Ofpe/WN9
\citep{1993ApJ...414..573T} and luminous blue variable stars
\citep{1997A&A...325..700N}. These young stars may indicate recent
star formation in the central region
\citep{1993ApJ...408..496M,2005astro.ph..7687N}, or they may have
migrated inward from larger distances to their current locations
\citep{2001ApJ...546L..39G}. In addition, the {\em Chandra} X-ray
Observatory has detected an unusually large number ($\apgt 2000$) of
hard X-ray (2--10 keV) point sources within 23\,pc of the Galactic
center \citep{2003ApJ...589..225M}. Seven of these sources are
transients, and are conjectured to contain stellar-mass black holes
\citep{2004astro.ph.12492M}; some may even harbor IMBHs
\citep{2001ApJ...558..535M}.

The Galactic center is a dynamic environment, where young stars and
star clusters form in molecular clouds \citep{2003ARA&A..41...57L} or
thick dusty rings \citep{2004astro.ph..9541N,2005astro.ph..7687N}, and
interact with their environment.  Several star clusters are known to
exist in this region \citep{1999ApJ...525..750F}, and the star
formation rate in the inner bulge is estimated to be comparable to
that in the solar neighborhood \citep{2001ApJ...546L.101P}, enough to
grow the entire bulge over the age of the Galaxy.

Of particular interest here are the several star clusters discovered
within $\sim 100$\,pc of the Galactic center, 11 of which have
reliable mass estimates \citep{2005A&A...435...95B}.  Most interesting
of these are the two dense and young ($\aplt 10$\,Myr) star clusters
Arches \citep{2002ApJ...581..258F} and the
Quintuplet\citep{1999ApJ...514..202F}, and the recently discovered
groups IRS\,13E \citep{2004A&A...423..155M} and IRS\,16SW
\citep{2005astro.ph..4276L}.

In this paper we study the relation between the star clusters in the
inner $\sim 100$\,pc of the Galactic center and, to some extend, the
partial formation of the central supermassive black hole.  In
particular we simulate the evolution of the star clusters born over a
range of distances from the Galactic center. While we follow their
internal dynamical evolution we allow the star clusters to spiral
inwards towards the Galactic center until they dissolve in the
background. During this process a runaway collision may have occurred
in the cluster and we follow the continuing spiral-in of the resulting
intermediate mass black hole.

Our prescription for building an intermediate mass black hole has been
well established in numerous papers concerning stellar collision
runaways in dense star clusters
\cite{1990ApJ...356..483Q,1999A&A...348..117P,2002ASPC..263..287F,
2004ApJ...604..632G,2005ApJ...628..236G, 2005astro.ph..3130F}. We just
build on these earlier results for our description of the collision
runaway and the way in which it leads to the formation of a black hole
of intermediate mass.

Eventually the IMBHs merge with the supermassive black hole, building
the SMBH in the process.  This model was initially proposed by
\citep{2001ApJ...562L..19E}, and here we validate the model by
detailed simulations of the dynamical evolution of individual star
clusters and the final spiral-in of the IMBH toward the SMBH. Using
the results of the direct N-body simulations we calibrate a
semi-analytic model to simulate a population of star clusters which
are born within $\sim 100$\,pc over the age of the Galaxy.

\section{Collision Runaways and Cluster Inspiral}\label{Sect:picture}

A substantial fraction of stars are born in clusters and these have a
power-law stellar mass functions fairly well described by a
``Salpeter'' exponent of -2.35, and with stellar masses ranging from
the hydrogen burning limit ($\sim 0.08$\,\msun) or a bit above
\citep{2005ApJ...628L.113S} to an upper limit of $\sim100\,\msun$ or
possibly as high as 150\,\msun\,\citep{2005Natur.434..192F}. The
massive stars start to sink to the cluster center immediately after
birth, driving the cluster into a state of core collapse on a time
scale $\tcc \simeq 0.2\trxh$
\citep{2002ApJ...576..899P,2004ApJ...604..632G}, where
\citep{1971ApJ...166..483S}
\begin{equation}
	\trxh \simeq 2\,{\rm Myr} \left( {r \over [{\rm pc}]} \right)^{3/2} 
	                     \left( {m \over [\msun]} \right)^{-1/2} 
	{n \over \Ll}.
\end{equation}
Here $m$ is the cluster mass, $r$ is its half-mass radius, $n$ is the
number of stars, and $\Ll \simeq \log(0.1n)\sim10$. In sufficiently
compact clusters the formation of a dense central subsystem of massive
stars may lead to a ``collision runaway,'' where multiple stellar
mergers result in the formation of an unusually massive object
\citep{1999A&A...348..117P,2002ApJ...576..899P,2004ApJ...604..632G,2005astro.ph..3130F}.
If the mass of this runaway grows beyond $\sim300$\,{\msun} it
collapses to an IMBH without losing significant mass in a supernova
explosion \citep{2003ApJ...591..288H}. Recently, this model has been
applied successfully to explain the ultraluminous X-ray source
associated with the star cluster MGG-11 in the starburst galaxy M82
\citep{2004Natur.428..724P}.  This model for creating an intermediate
mass black hole in a dense star cluster was adopted by
\cite{2005ApJ...628..236G}, who continued by studying the evolution of
massive $\apgt 10^6$\,\msun\, star clusters within about 60\,pc from
the Galactic center. Their conclusions are consistent with the earlier
$N$-body models
\cite{2000ApJ...545..301K,2003ApJ...593..352P,2003ApJ...596..314M,2004ApJ...607L.123K}
and analytic calculations \cite{2001ApJ...546L..39G} in that massive
clusters can reach the galactic center but in doing so they populate
the inner few parsecs with a disproportionately large number of massive
stars.

The main requirement for a successful collision runaway is that the
star cluster must experience core collapse (i) before the most massive
stars explode as supernovae ($\sim3$\,Myr) and (ii) before the cluster
dissolves in the Galactic tidal field.  The collisional growth rate
slows dramatically once the runaway collapses to an IMBH. We estimate
the maximum runaway mass achievable by this process as follows.  For
compact clusters ($\trxh\aplt100$ Myr), essentially all the massive
stars reach the cluster core during the course of the runaway, and the
runaway mass scales with the cluster mass:
$\mrunaway\simeq8\times10^{-4}m\,\Ll$ \citep{2002ApJ...576..899P}. For
systems with longer relaxation times, only a fraction of the massive
stars reach the core in time and the runaway mass scales as
$m\trxh^{-1/2}$ \citep{2004astro.ph.12622M} (see their Eq. 11).  The
relaxation based argument may result in higher mass runaways in star
clusters with a very small relaxation time compared to the regime
studied in Monte Carlo N-body simulations \citep{2004astro.ph.12622M}.
A convenient fitting formula combining these scalings, calibrated by
N-body simulations for Salpeter-like mass functions, is
\citep{2002ApJ...576..899P,2004astro.ph.12622M}
\begin{equation}
    \mrunaway \sim 0.01 m
	      \left( 1 + \frac{\trxh}{\rm 100 Myr} \right)^{-1/2}\,.
\end{equation}
Early dissolution of the cluster reduces the runaway mass by
prematurely terminating the collision process.

As core collapse proceeds, the orbit of the cluster decays by
dynamical friction with the stars comprising the nuclear bulge.  The
decay of a circular cluster orbit of radius $R$ is described by (see
[Eq.~7-25] in \cite{1987gady.book.....B}, or
\cite{2003ApJ...596..314M} for the more general case):
\begin{equation}
	\frac{dR}{dt} = -0.43 {Gm \LL \over R^{(\alpha + 1)/2} v_c}\,,
\label{Eq:df}
\end{equation}
where $v_c^2 = GM(R)/R$, $\alpha = 1.2$, $M(R)$ is the mass within a
distance $R$ from the Galactic center and we take $\LL\sim8$
\citep{2003MNRAS.344...22S}. Numerical solution of this equation is
required due to the complicating effects of stellar mass loss, which
drives an adiabatic expansion of the cluster, and by tidal stripping,
whereby the cluster mass tends to decrease with time according to $m(t)
= m_0(1 - \tau/\tdis)$, \citep{2002ApJ...576..899P}.
Here $m_0$ is the initial mass of the cluster, $\tau$ is the cluster
age in terms of the instantaneous relaxation time (\trxJ) within the
Jacobi radius, and $\tdis$ is the time scale for cluster disruption:
$\tdis \simeq 0.29\trxJ$ \footnote{Theoretical considerations suggest
that the time scale for cluster dissolution has the form $\tdis = k
\thc^{1/4} \trxh^{3/4}$, where {\thc} is the cluster crossing time
\citep{2003MNRAS.340..227B}. The constant $k$ may be obtained from
direct N-body simulations of star clusters near the Galactic center
\cite{2001ApJ...546L.101P}, resulting in $k\simeq 7.5$, with {\thc}
and {\trxh} expressed in Myr.}.

Even after the bulk of the cluster has dissolved, a dense stellar cusp
remains surrounding the newborn IMBH, and accompanies it on its
descent toward the Galactic center.  The total mass of
stars in the cusp is typically comparable to that of the IMBH itself
\citep{2004ApJ...613.1143B} and it is composed predominantly of
massive stars, survivors of the population that initiated the core
collapse during which the IMBH formed.  Eventually even that cusp
slowly decays by two-body relaxation \citep{2003ApJ...593L..77H},
depositing a disproportionately large number of massive stars and the
orphaned IMBH close to the Galactic center \cite{2005ApJ...628..236G}.
Ultimately, the IMBH merges with the SMBH.

\section{Simulating star clusters within $\sim 100$\,pc 
	 from the Galactic center}
\label{Sect:Nbody}

We have performed extensive direct N-body calculations to test the
validity of the general scenario presented above, and to calibrate the
semi-analytic model.  Our analysis combines several complementary
numerical, analytical and theoretical techniques in a qualitative
model for the formation and evolution of the nuclear bulge of the
Milky Way Galaxy.  The semi-analytical model outlined in
Sect.\,\ref{Sect:picture}, and which is based on equation \ref{Eq:df}
of \cite{2003ApJ...596..314M}, is based on simple characterizations of
physical processes, which we calibrate using large-scale
N-body simulations.  The initial conditions for these simulations are
selected to test key areas in the parameter space for producing IMBHs
in the inner $\sim 100$\,pc of the Galactic center.

The N-body calculations employ direct integration of Newton's
equations of motion, while accounting for complications such as
dynamical friction and tidal effects due to the Galactic field,
stellar and binary evolution, physical stellar sizes and the
possibility of collisions, and the presence of a supermassive black
hole in the Galactic center.  Two independent but conceptually similar
programs are used: (1) the ``kira'' integrator, part of the Starlab
software environment (see {\tt
http://www.manybody.org/$\sim$manybody/starlab.html},
\cite{2001MNRAS.321..199P}), and (2) NBODY4 \\ (see {\tt
http://www.sverre.org}) \citep{Aarseth2003}. Both codes achieve their
greatest speed, as in the simulations reported here, when run in
conjunction with the special-purpose GRAPE-6 (see {\tt
http://www.astrogrape.org}) hardware acceleration
\citep{2003PASJ...55.1163M}.

Both kira and NBODY4 incorporate parametrized treatments of stellar
evolution and allow for the possibility of direct physical collisions
between stars, thus including the two key physical elements in the
runaway scenario described here (see also
\cite*{2004Natur.428..724P}). A collision occurs if the distance
between two stars becomes smaller than the sum of the stellar radii,
except that, for collisions involving black holes, we use the tidal
radius instead.  During a collision mass and momentum are conserved.
These are reasonable assumptions since the relative velocity of any
two colliding stars is typically much smaller than the escape speed
from either stellar surface
\citep{2003MNRAS.345..762L,2005MNRAS.358.1133F}.

We performed N-body simulations of star clusters containing up to
131,072 stars and starting at $R=1$\,pc, 2, 4, 10 and 100\,pc from the
Galactic center, with various initial concentrations ($W_0=6$ and 9)
and with lower limits to the initial mass function of 0.1\,\msun\, and
1\,\msun. These simulations were carried out as part of the
calibration of the semi-analytic model which we presented in
Sect.\,\ref{Sect:analytic}.

\begin{figure}
\epsscale{.7}\plotone{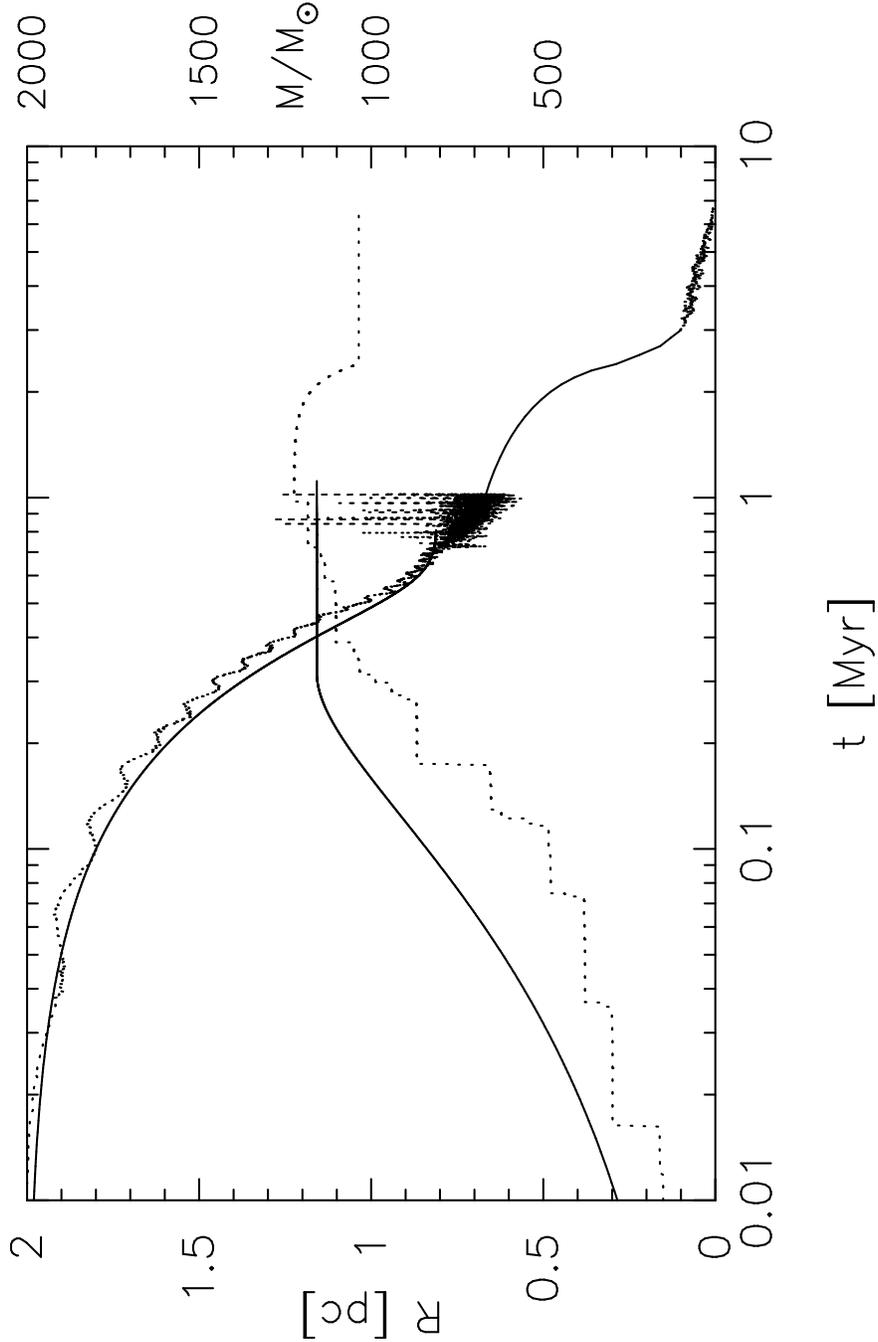}
\caption[]{
Orbital evolution of a star cluster.  A 45,000\,{\msun} cluster of
65,536~stars with a Salpeter initial mass function, a lower mass limit
of 0.2\,{\msun}, and a $W_0=9$ King model initial density profile
spirals in to the Galactic center from an distance of 2\,pc (lines
from top left to bottom right), while producing an IMBH via collision
runaway (bottom left to top right; scale on the right axis).  Solid
lines (based on equation \ref{Eq:df}) show the results of the
semi-analytic model (with $\log\Lambda=8$), the dotted lines represent
high-precision N-body calculations.  The solid and dotted lines match
quite well, indicating that the analytic model produces satisfactory
results. 
\label{fig:N-body}	
}
\end{figure}

One such comparison is presented in Figure\,\ref{fig:N-body}, which
shows the orbital evolution and runaway growth in a star cluster born
with 65536 stars in a circular orbit at a distance of 2\,pc from the
Galactic center.  The solid lines in the figure result from the
semi-analytic model (based on equation \ref{Eq:df} and
\cite{2003ApJ...596..314M}), while the high precision N-body
calculations are represented by dotted lines. They match quite well,
indicating that the simple analytic model produces satisfactory
results in reproducing the general features and physical scales of the
evolution.

\begin{figure}
\epsscale{1}\plotone{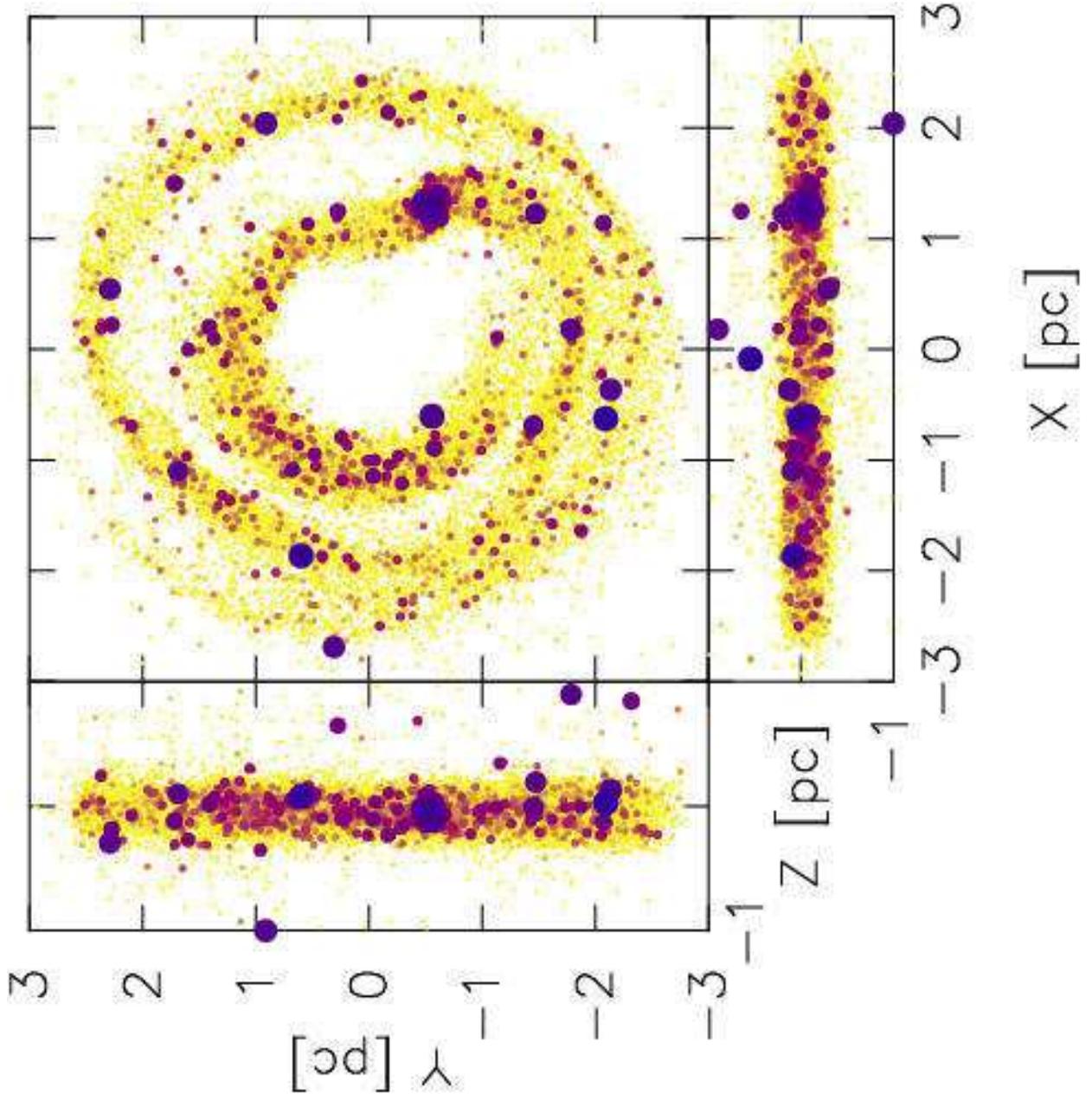}
\caption[]{ 
Snapshot of a dissolving cluster, showing the
projection on various planes of the N-body simulation shown in
Figure\,\ref{fig:N-body}, at a cluster age of 0.35\,Myr.  The cluster
initially orbited in the $X$--$Y$-plane in a circular orbit of radius
of 2 pc from the Galactic center. By the time shown the cluster has
lost about 30\% of its initial mass.  However, the majority of massive
stars are still bound to the IMBH progenitor as it orbits the Galactic
center at a distance of roughly 1.2\,pc.  The individual stars shown
range in mass from 0.2\,{\msun} to 100\,{\msun}, with symbol sizes
proportional to the surface area of each star.  Color runs from red
(cool) to blue (hot).
\label{fig:image}
}
\end{figure}

As the cluster in Figure\,\ref{fig:N-body} sinks toward the Galactic
center, it produce one massive star through the collision runaway
process.  In Figure\,\ref{fig:image} we show a snapshot of this
simulations projected in three different planes at an age of
0.35\,Myr.  By this time $\sim 30\%$ of the cluster has already
dispersed and its stars have spread out into the shape of a disk
spanning the inward-spiraling orbit.  By the time of Figure
\ref{fig:image}, a $\sim 1100$\,{\msun} collision runaway star has
formed in the cluster center; this object subsequently
continues to grow by repeated stellar collisions.  The growth of the
collision runaway is indicated by the dotted line in
Figure\,\ref{fig:N-body} running from bottom left to top right (scale
on the right vertical axis).

By an age of about 0.7\,Myr the cluster is almost completely disrupted
and the runaway process terminates.  After the cluster dissolves, the
IMBH continues to sink toward the Galactic center, still accompanied
by 10--100 stars which initially were among the most massive in the
cluster.

Near the end of its lifetime, the runaway star loses about
200\,{\msun} in a stellar wind and subsequently collapses to a $\sim
1000$\,\msun\, IMBH at about 2.4\,Myr.  The IMBH and its remaining
stellar companions continue to sink toward the Galactic center.  The
continuing ``noise'' in the dotted curve in Figure \ref{fig:N-body}
reflects the substantial eccentricity of the IMBH orbit.  At an age of
2.5--3\,Myr, the remnant star cluster consisting of an IMBH orbited by
a few of the most massive stars, quite similar to the observed star
cluster IRS\,13, arrives in the inner 0.1 pc of the Galaxy (see
sect.\,\ref{Sect:Observations}).

\section{Merger with the central black hole}\label{Sect:finalparsec}

When the IMBH arrives within about 0.1\,pc of the Galactic center the
standard formula for dynamical friction\,\citep{1987gady.book.....B} is
becoming unreliable, as the background velocity dispersion increases
and the effects of individual encounters become more significant.  It
is important, however, to ascertain whether the IMBH spirals all the
way into the SMBH, or if it stalls in the last tenth of a parsec, as
higher-mass black holes may tend to do \citep{2005ApJ...621L.101M}.

To determine the time required for the IMBH to reach the central SMBH,
we have performed additional N-body calculations, beginning with a
1000\,{\msun} and a 3000\,{\msun} IMBH in circular orbits at a
distance of 0.1\,pc from the Galactic center.  Both IMBHs are assumed
to have shed their parent cluster by the start of the simulation.  The
inner parsec of the Galaxy is modeled by 131,071 identical stars with
a total stellar mass of $4\times 10^6$\,{\msun}, distributed according
to a $R^{-1.75}$ density profile; a black hole of $3\times
10^6$\,{\msun} resides at the center.  The region within a milliparsec
of the central SMBH is depleted of stars in our initial
conditions. This is supported by the fact that the total Galactic mass
inside that radius, excluding the central SMBH is probably less than
$10^3\,\msun$\, \cite{1998ApJ...509..678G,2003ApJ...594..812G}. We
stop the calculations as soon as the IMBH reaches this distance.

\begin{figure}
\epsscale{.7}\plotone{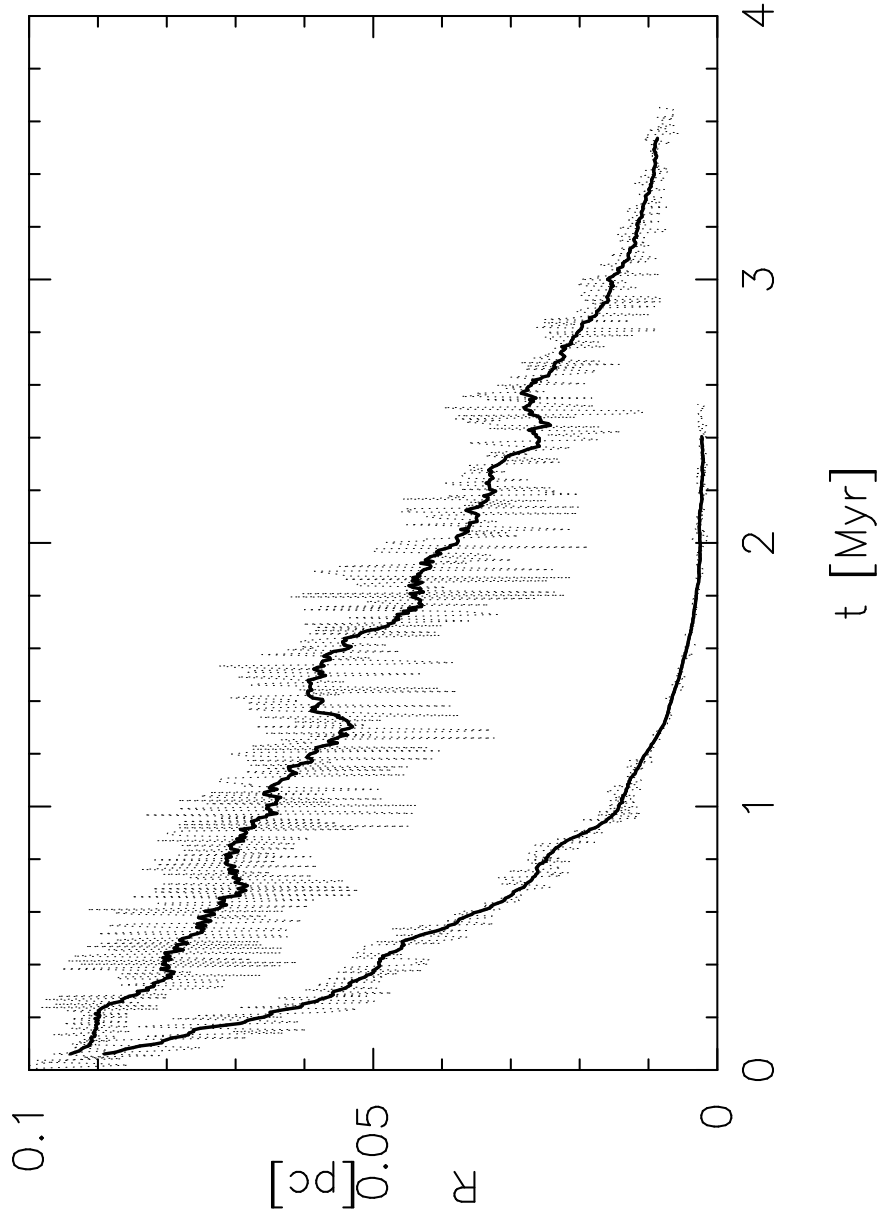}
\caption[]{ 
Final orbital evolution of two IMBHs, of masses
1000\,{\msun} (right curve) and 3000\,{\msun} (left curve), each
starting from a circular orbit of radius 0.1\,pc and ending at the
center of the Galaxy, where the IMBHs ultimately merge with the SMBH
there.  The dotted lines show the actual orbits, while the solid lines
have been smoothed over several orbits to filter out the
short-timescale fluctuations due to orbital eccentricity.  The final
merger occurs on a timescale similar to the time interval between
successive IMBH arrivals in the central parsec, driving the growth by
accretion of the central black hole.
\label{fig:final_parsec}
}
\end{figure}

Figure\,\ref{fig:final_parsec} (see also the dotted line in
Figs.\,\ref{fig:N-body}) shows the orbital evolution of the
1000\,{\msun} and 3000\,{\msun} IMBHs in our simulations.  Although
the black-hole orbits are initially circular, eccentricities on the
order of $\aplt 0.6$ are induced quite quickly by close encounters
with field stars.  The rate of spiral-in near the SMBH is smaller than
farther out, because the increasing velocity dispersion tends to
reduce the effect of dynamical friction and because the IMBH reaches
the inner depleted area.

The central milliparsec was initially empty in our simulations, and
there was insufficient time to replenish it during our calculations.
It is unlikely that sufficient stellar mass exists within this region
for dynamical friction to drive the IMBH much closer to the
SMBH. (Interestingly, this distance is comparable to the orbital
semi-major axis of the star S0-2, which is observed in a 15 year orbit
around the Galactic center \cite{2003ApJ...586L.127G}.) The time scale
for a 1 mpc orbit to decay by gravitational radiation exceeds the age
of the Galaxy for circular motion, so unless the IMBH orbit is already
significantly eccentric, or is later perturbed to higher eccentricity
($\apgt 0.9$ to reduce the merger time to $\aplt 10^9$\,years) by
encounters with field stars or another IMBH, the orbital decay
effectively stops near the central SMBH.

    While the IMBH stalls, another star cluster may form, sink toward
    the Galactic center, and give rise to a new IMBH which
    subsequently arrives in the inner milliparsec (see Sec. 5).  This
    process will be repeated for each new IMBH formed, until
    interactions become frequent enough to drive a flux of IMBHs into
    the loss cone where gravitational radiation can complete the
    merger process.

We can estimate the number of IMBHs in a steady state in the inner
few milliparsecs of the SMBH. The time scale for a close (90 degree
deflection) encounter in a system of $n_{\rm IMBH}$ IMBHs is
\begin{equation}
	t_{\rm close} \sim \left(\frac{M_{\rm SMBH}}{m_{\rm IMBH}}\right)^2
		           \frac{t_{\rm orb}}{n_{\rm IMBH}}\,,
\end{equation}
where $M_{\rm SMBH}$ and $m_{\rm IMBH}$ are the masses of the SMBH and
the IMBH, respectively, and $t_{\rm orb} \sim 1-10$ years is the
typical orbital period at a distance of 1 mpc from the SMBH.  For
$M_{\rm SMBH} \sim 10^6\msun$ and $m_{\rm IMBH} \sim 10^3\msun$, we
find $t_{\rm close} \sim 1-10 \times 10^6/n_{\rm IMBH}$ years,
comparable to the in-fall time scale unless $n_{\rm IMBH}$ is large.

Close encounters are unlikely to eject IMBHs from the vicinity of the
Galactic center, but they do drive the merger rate by replenishing the
loss cone around the
SMBH \citep{2004ApJ...606..788M,2004ApJ...616..221G}. As IMBHs
accumulate, the cusp around the SMBH eventually reaches a steady state
in which the merger rate equals the rate of in-fall, with a roughly
constant population of a few IMBHs within about a milliparsec of the
SMBH. A comparable analysis was performed by
\citep{2004ApJ...606L..21A} for stellar mass black holes around the
SMBH, and if we scale their results to IMBHs we arrive at a similar
steady state population.

\section{The evolution of a population of star clusters}
\label{Sect:analytic}

We now turn to the overall evolution of the population of clusters
which gave rise to the nuclear bulge.  We have performed a Monte-Carlo
study of the cluster population, adopting a star formation rate that
declines as $1/t$ over the past 10 Gyr
\citep{2004Natur.428..625H}. Cluster formation times are selected
randomly following this star formation history, and masses are
assigned as described below, until the total mass equals the current
mass of the nuclear bulge within 100 pc of the Galactic center---about
$10^9$\,\msun.  The total number of clusters thus formed is $\sim
10^5$ over the 10 Gyr period.

For each cluster, we select a mass ($m$) randomly from a cluster
initial mass function which is assumed to follow the mass distribution
observed in starburst galaxies---a power-law of the form $N(m) \propto
m^{-2}$ between $\sim10^3$\,{\msun} and $\sim10^7$\,{\msun}
\citep{1999ApJ...527L..81Z}. The distance to the Galactic center ($R$)
is again selected randomly, assuming that the radial distribution of
clusters follows the current stellar density profile in the bulge
between 1\,pc and
100\,pc\,\citep{1997MNRAS.284..576E,1998ApJ...509..678G}.  The current
distribution of stars must reflect the formation distribution to a
large extent, because most stars' orbits don't evolve significantly,
but only the orbits of the more massive stellar clusters.
The initial density profiles of the clusters are assumed to be
$W_0=6$--9 King models. This choice of high-concentrated King models
is supported by the recent theoretical understanding by
\cite{2004ApJ...608L..25M} of the relation between age and core-radius
for young star clusters in the large Magellanic cloud observed by
\cite{2003MNRAS.338...85M}.

We establish a cluster mass-radius relation by further assuming that
clusters are born precisely filling their Jacobi surfaces in the
Galactic tidal field.  This provides a lower limit to the fraction of
clusters that produce an IMBH and sink to the Galactic center.

The evolution of each cluster, including specifically the moment at
which it undergoes core collapse, the mass of the collision runaway
(if any) produced, and the distance from the Galactic center at which
the cluster dissolves, is then calculated deterministically using our
semi-analytic model.  After cluster disruption, the IMBH continues to
sink by dynamical friction, eventually reaching the Galactic center.

\subsection{Results of the cluster population model}\label{Sect:Results}

\begin{figure}
\epsscale{.5}\plotone{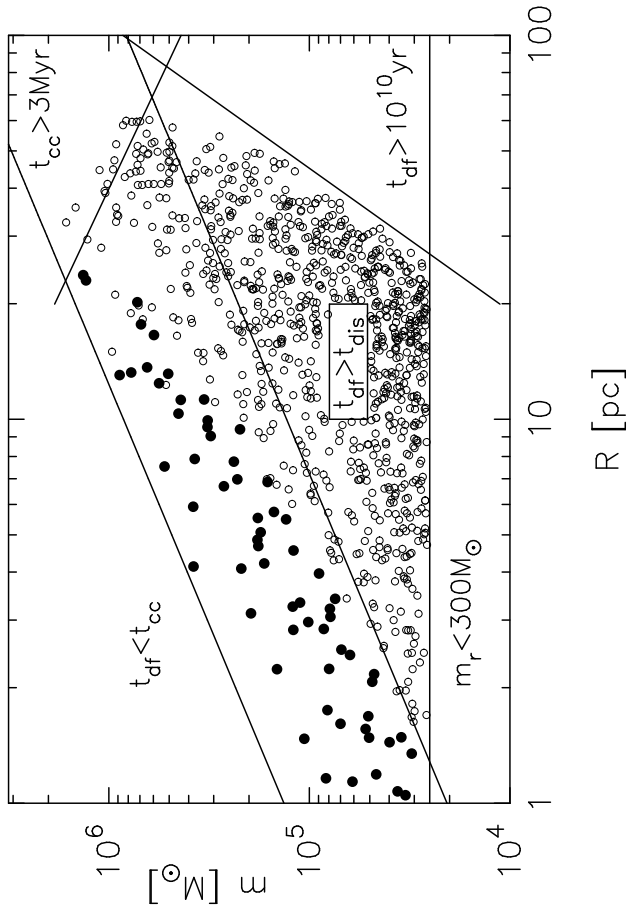}
\caption[]{
Parameter space for the formation of IMBHs.  Within
the semi-analytic model, the initial mass of a star cluster ($m$) and
its initial distance from the Galactic center ($R$) determine its
fate, and control whether or not an IMBH forms.  For the initial mass
function for the cluster stars, we choose a Salpeter distribution
between 0.2 and 100\,{\msun} and King $W_0=9$.  The lines separate the
various regions of the parameter space investigated.  Star clusters
born in the top left part of the diagram ($\tdf<\tcc$) spiral inward
and dissolve in the Galactic field before core collapse can occur.  At
the top right ($\tcc>3$\,Myr), the most massive stars in the cluster
leave the main sequence before core collapse has occurred, thus
preventing a collision runaway.  Clusters born with masses less than a
few $10^4$\,{\msun} cannot form a sufficiently massive collision
product ($\mrunaway<300$\,\msun), and clusters to the right and below
the middle solid curve ($\tdf>\tdis$) dissolve before they reach the
central parsec.  To the right of and below the rightmost diagonal line
($\tdf>10^{10}$\,year), there is insufficient time for IMBHs to form
and sink to the center of the Galaxy.  Open and filled circles
represent initial conditions that result in an IMBH reaching the
central parsec by the present day. Filled circles indicate that part
of the parent star cluster is still present upon arrival.  Open
circles represent cases where the IMBH continued to sink to the
Galactic center even after its parent star cluster dissolved,
typically at a rather large distance from the Galactic center.  The
two sets of symbols are roughly separated by the line for which the
disruption time equals the dynamical friction time scale: $t_{\rm
df}=t_{\rm dis}$.
\label{fig:MC}		
}
\end{figure}

Figure\,\ref{fig:MC} summarizes how the fates of the star clusters in
our simulation depend on $m$ and $R$.  Open and filled circles
represent initial conditions that result in an IMBH reaching the
central parsec by the present day.  The various lines define the
region of parameter space expected to contribute to the population of
IMBHs within the central parsec, as described in the caption.  Here we
emphasize that our results depend linearly on the fraction of stars in
the bulge that form in star clusters. The number of star clusters and
IMBHs is proportional to this factor, which is not necessarily
constant with time.  Bear in mind also that, though theoretical
uncertainties are about a factor of two, the systematic uncertainties
can be much larger and depend critically on various assumptions in the
models, like the amount of mass loss in the stellar wind of the
collision product and the fate of the stellar remnant in the supernova
explaion. The results of our calculations may be summarized as
follows:
 
\begin{enumerate}

\item 5\% -- 10\% of star clusters born within 100\,pc of the Galactic
center produce an IMBH.

\item The mean mass of IMBHs now found in the inner 10\,pc is $\sim
1000$\,\msun, whereas IMBHs between 90 and 100\,pc average $\sim
500$\,\msun.

\item Over the age of the Galaxy ($\sim10^{10}$ years) a total of
1000--3000 IMBHs have reached the Galactic center, carrying a total
mass of $\sim 1\times10^6\msun$. Here the range in masses stems from
variations in the adopted stellar mass function.

\item At any instant, approximately $\sim 50$ IMBHs reside in the
inner 10\,pc, about ten times that number lie within the nuclear star
cluster (inner 30\,pc), and several lie within the innermost few
tenths of a parsec.

\item One in every $\sim 30$ IMBHs is still accompanied by a remnant
of its young (turn-off mass $\apgt 10$\,\msun) star cluster when it
reaches the inner parsec, resulting in a few IMBHs at any time in the
inner few parsecs with young stars still bound to them, much like
IRS\,13E or IRS\,16SW.

\end{enumerate}

On the basis of our N-body simulations of the central 0.1 pc in
Sect.\,\ref{Sect:finalparsec} we expect that the majority of IMBHs
which arrive in the Galactic center eventually merge with the SMBH on
a time scale of a few Myr, driven by the emission of gravitational
radiation and interactions with local field stars and other IMBHs.  In
our simulations the in-fall rate has increased over the lifetime of the
Galaxy (following our assumed star formation rate), from one arrival
per $\sim 20$\,Myr to the current value of one every $\sim 5$\,Myr,
with a time average IMBH in-fall rate of roughly one per $\sim 7$\,Myr.
(A lower minimum mass in the initial mass function produces higher
in-fall rates.)

Some of the field stars near the SMBH may be ejected from the Galactic
center with velocities of up to $\sim 2000$\,km/s following encounters
with the hard binary system formed by the IMBH and the central SMBH
\citep{1988Natur.331..687H,2003ApJ...599.1129Y,2005MNRAS.363..223G}. Support for this
possibility comes from the recent discovery of SSDS\,J090745.0+024507,
a B9 star with a measured velocity of 709\,km/s directly away from the
Galactic center \citep{2005ApJ...622L..33B}.

IMBHs are potentially important sources of gravitational wave
radiation.  A merger between a 1000\,{\msun} IMBH and a $\sim 3 \times
10^6$\,{\msun} SMBH would be detectable by the LISA gravitational wave
detector to a distance of several billion parsecs. Assuming that the
processes just described operate in most spiral galaxies, which have a
density of roughly 0.002 Mpc$^{-3}$\, \cite{2004MNRAS.353..713K}, we
estimate a detectable IMBH merger rate of around two per week, with a
signal to noise $\sim 10^3$.

\subsection{The current cluster population}
\label{Sect:currentclusters}

\begin{figure}
\epsscale{.7}\plotone{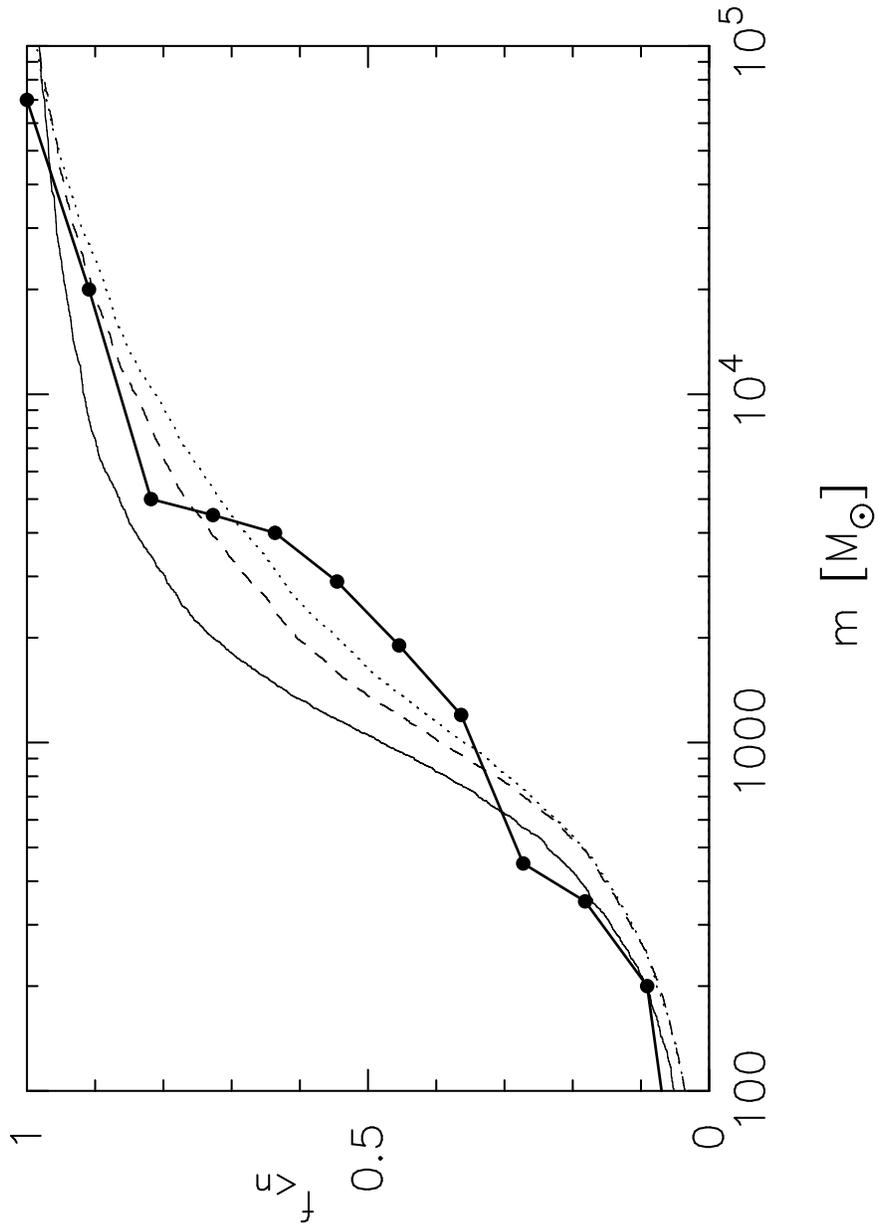}
\caption[]{ 
Cumulative distribution of star clusters in the
vicinity of the Galactic center. The thick solid curve gives the
masses of eleven observed Galactic center clusters (bullets) with
reliable mass estimates
\citep{2005A&A...435...95B,1999ApJ...525..750F}.  The other curves
give the distribution of cluster masses from our semi-analytic
model. The theoretical curves are quite insensitive to the star
formation history, but change with the lower mass limit adopted for
the cluster initial mass function. The lower limit is 1\,{\msun} for
the thin solid curve, 0.2\,{\msun} for the dashed curve, and
0.1\,{\msun} for the dotted curve.  The number of observable clusters
is best matched with the decaying star formation rate adopted in
Sect.\,\ref{Sect:analytic}.
\label{fig:Borissova}
}
\end{figure}

Our semi-analytic model for the evolution of star clusters in the
inner $\sim 100$\,pc of the Galaxy yields a steady-state distribution
of cluster masses which we can compare with observed star clusters in
the vicinity of the Galactic center.  Figure\,\ref{fig:Borissova}
compares the observed mass distribution of young star clusters in the
bulge with our steady-state solution. The data include the Arches
cluster \citep{2002ApJ...581..258F}, the Quintuplet
\citep{1999ApJ...514..202F}, IRS\,13E \citep{2004A&A...423..155M},
IRS\,16SW \citep{2005astro.ph..4276L}, and 7 recently discovered star
clusters with reliable mass estimates \citep{2005A&A...435...95B}. For
comparison we show a realization of the present-day population of star
cluster masses generated by our semi-analytic model.

Using the adopted declining star-formation rate from
Sect.\,\ref{Sect:analytic} (see \cite{2004Natur.428..625H}), we find
about $\sim 50$ star clusters within the central 100 pc at any given
time, consistent with the earlier prediction of
\cite{2001ApJ...546L.101P}. Assuming a flat (i.e.~uniform) star
formation rate, we predict $\sim 400$ clusters in the same region,
about an order of magnitude more than currently observed.  In our
semi-analytic model, about 15\% of all present-day star clusters host
an IMBH or are in the process of producing one.  Between 1\% and 8\%
of star clusters with a present-day mass less than $10^4$\,{\msun}
contain an IMBH, whereas more than 80\% of clusters with masses
between 30,000 and $2\times 10^5$\,{\msun} host an IMBH.  For more
massive clusters the probability of forming an IMBH drops sharply.

Finally, we note that we are rather unlikely to find an orphaned very
massive ($\apgt 200$\,\msun) star.  During the last 1 Gyr, only about
10--40 of such objects have formed in the inner 100\,pc of the
Galaxy.  Lower mass merger products, however, are quite common.  The Pistol
star\citep{1999ApJ...514..202F} may be one observational example.

\section{Discussion}

\subsection{Evolution of the merger product}\label{Sect:massloss}

One of the main uncertainties in our calculations is whether or not
mass gain by stellar collisions exceeds mass loss by stellar winds.
Although the accretion rate in our models is very high, mass loss
rates in massive stars are uncertain, and it is conceivable that
sufficiently high mass loss rates might prevent the merger product
from achieving a mass of more than a few hundred \msun.

Mass loss in massive ($\apgt 100$\,\msun) stars may be radiatively
driven by optically thin lines. In this case it is possible to derive
upper limits to the mass loss. Such calculations, including the von
Zeipel \citep{1924MNRAS..84..665V} effect for stars close to the
Eddington-Gamma limit, indicate that stellar wind mass loss rates may
approach $10^{-3}$\,\msun\,yr$^{-1}$ \citep{2000A&A...362..295V}. If the
star is rotating near the critical rate, the mass loss rate may be
even larger \citep{2004A&A...418..639A}. Outflow velocities, however,
may be so small that part of the material falls back on the equatorial
zone, where the mass loss is least \citep{2004A&A...418..639A}. The
calculations of \citep{2004A&A...418..639A} match the observed mass
loss rates for $\eta$ Carina, which has a peak of $1.6\pm
0.3\times10^{-3}$\,\msun\,yr$^{-1}$ (assuming spherical symmetry)
during normal outbursts, falling to $10^{-5}$\,\msun\,yr$^{-1}$ during
the intervening 5.5\,years \citep{2003A&A...410L..37V}. For young
($\aplt4$\,Myr) O stars in the small Magellanic cloud low ($\aplt
10^{-8}$\,\msun/yr) mass loss rates were
observed \citep{2004A&A...420.1087M}, indicating that massive stars may
have much lower mass loss rates until they approach the end of their
main-sequence lifetimes (see \cite{2003A&A...404..975M}).

Thus it remains unclear whether the periods of high mass loss persist
for long enough to seriously undermine the runaway scenario adopted
here.  We note that the collision runaways in our simulations are
initiated by the arrival of a massive star in the cluster core
\citep{1999A&A...348..117P}. If such a star grows to exceed $\sim
300$\,{\msun}, most collisions occur within the first 1.5\,Myr of the
cluster evolution.  The collision rate during the period of rapid
growth typically exceeds one per $\sim 10^4$ years, sustained over
about 1 Myr, resulting in an average mass accretion rate exceeding
$10^{-3}$\,\msun/yr, comparable to, and possibly exceeding, the
maximum mass loss rates derived for massive stars.  Furthermore, in
our N-body simulations (and in the semi-analytic model), the stellar
mass loss rate increases with time, with little mass loss at the
zero-age main sequence and substantially more near the end of the
main-sequence stellar lifetime ($\dot{m}_{\rm wind} \propto L^{2.2}$)
\citep{2000A&A...362..295V,1994A&A...290..819L,2002ApJ...577..389K}.
In other words, mass loss rates are relatively low while most of the
accretion is occurring.  This prescription for the mass loss rate
matches that of evolutionary calculations for massive Wolf-Rayet stars
\citep{2005A&A...429..581M}.

We also emphasize that a large mass loss rate in the merger product
{\em cannot} prevent the basic mass segregation and collision process,
even though it might significantly reduce the final growth rate
\citep{1999A&A...348..117P}.  These findings are consistent with
recent N-body simulations of small clusters in which the assumed mass
loss rate from massive ($>120$\,\msun) stars exceeded
$10^{-3}$\msun/yr \citep{2004astro.ph.10510B}.

The stellar evolution of a runaway merger product has never been
calculated in detail, and is poorly understood.  However, it is worth
mentioning that its thermal time scale significantly exceeds the mean
time between collisions.  Even if the star grows to $\apgt
10^3$\,{\msun}, the thermal time scale will be 1--$4\times
10^4$\,years, still comparable to the collision rate.  The accreting
object will therefore be unable to reestablish thermal equilibrium
before the next collision occurs.

We note in passing that the supermassive star produced in the runaway
collision may be hard to identify by photometry if the cluster
containing it cannot be resolved: The runaway is mainly driven by
collisions between massive stars, which themselves have luminosities
close to the Eddington-Gamma limit. Since the Eddington luminosity
scales linearly with mass, a collection of luminous blue variables at
the Eddington luminosity are comparable in brightness to an equally
massive single star. Spectroscopically, however, the collision runaway
may be very different.

Mass loss in the form of a dense stellar wind before the supernova can
dramatically reduce the mass of the final black hole, or could even
prevent black hole formation altogether \citep{2003ApJ...591..288H}.
The runaway merger in fig.\,\ref{fig:N-body} develops a strong stellar
wind near the end of its lifetime before collapsing to a
$\sim1000$\,{\msun} IMBH at $\sim2.4$\,Myr.  It is difficult to
quantify the effect of stellar winds on the final IMBH mass because
the mass loss rate of such a massive star remains uncertain
\citep{2001A&A...369..574V}.  However, it is important to underscore
here the qualitative results that stellar winds are unable to prevent
the occurrence of repeated collisions, and significantly limit the
outcome only if the mass loss rate is very high---more than
$\sim10^{-3}\,\msun/{\rm yr}$---and sustained over the lifetime of the
star.

\section{The star clusters IRS13E and IRS16SW}
\label{Sect:Observations}

The best IMBH candidate in the milky-way Galaxy was recently
identified in the young association IRS\,13E in the Galactic center
region.  IRS\,13E is a small cluster of stars containing three
spectral type O5I to O5III and four Wolf-Rayet stars, totaling at most
$\sim300\,\msun$ \citep{2004A&A...423..155M,2005ApJ...625L.111S}. (The
recently discovered cluster IRS\,16SW\citep{2005astro.ph..4276L} also
lies near the Galactic center and reveals similarly interesting
stellar properties.)  Both clusters are part of the population of
helium-rich bright stars in the inner parsec of the Galactic center
\citep{2001A&A...366..466P}. With a ``normal'' stellar mass function,
as found elsewhere in the Galaxy, stars as massive as those in
IRS\,13E are extremely rare, occurring only once in every $\sim 2000$
stars.  However, in the Galactic center, a ``top-heavy'' mass function
may be common \citep{2004astro.ph..9415F,2005ApJ...628L.113S}.

The mean proper motion of five stars in IRS\,13E is $\langle
v\rangle_{\rm 2D} = 245$\,km/s;\citep{2005ApJ...625L.111S} an
independent measurement of four of these stars yields $270$\,km/s
\citep{2004A&A...423..155M}.  If IRS13E were part of the rotating
central stellar disk \citep{2003ApJ...594..812G}, this would place the
cluster $\sim 0.12$\,pc behind the plane on the sky containing the
SMBH, increasing its galactocentric distance to about 0.18\,pc,
consistent with a circular orbit around the SMBH at the observed
velocity.  The five IRS\,16SW stars have $\langle v\rangle_{\rm 2D}
\simeq 205$\,km/s \citep{2005astro.ph..4276L}, corresponding to the
circular orbit speed at a somewhat larger distance ($\sim 0.4$\,pc).

The greatest distance between any two of the five stars in IRS\,13E
with known velocities is $\sim 0.5$\,seconds of arc (0.02\,pc at
8.5\,kpc), \citep{2004A&A...423..155M,2005ApJ...625L.111S} providing a
lower limit on the Jacobi radius: $\rJ \apgt 0.01$\,pc.  It then
follows from the Hills equation ($\rJ^3 \simeq R^3 m/M$) that the
minimum mass required to keep the stars in IRS\,13E bound is about
1300\,{\msun} (see also \citep{2004A&A...423..155M}).

A more realistic estimate is obtained by using the measured velocities
of the observed stars, using the expression: $m = \langle v^2 \rangle
R/G$. The velocity dispersion of all stars, E1, E2, E3, E4, and E6, is
about $\langle v \rangle \simeq 68$--84km/s, which results in a
estimated mass mass of about 11000--16000\,\msun.  Such a high mass
would be hard to explain with the collision runaway scenario.

However, the stars E1 and E6 may not be members. The extinction of the
latter star is smaller than that of the other stars, indicating that
it may be closer to the sun than the rest of the cluster and therefore
not a member \citep{2005ApJ...625L.111S}.  One could also argue that
star E1 should be excluded from the sample.  With a high velocity in
the opposite direction of the other stars it is equally curious as
star E6 in both velocity space and the projected cluster image, where
it is somewhat off from the main cluster position.  Without star E1
the velocity dispersion of the cluster becomes $\langle v \rangle
\simeq 47$--$50$km/s, which results in a estimated mass mass of about
5100--5800\,\msun.

These estimates for the total cluster mass are upper limits for the
estimated mass of the dark point mass in the cluster center.  If the
cluster potential is dominated by a point mass object with a total
mass exceeding the stellar mass by a seizable fraction, the stars are
in orbit around this mass point. In that case some of the stars may be
near the pericenter of their orbit. Since the velocity of a star at
pericenter will be a factor of $\sim \sqrt{2}$ larger compared to the
velocity in a circular orbit, the estimated black hole mass may
therefore also be smaller by up to a factor of 2.

We stress that the IMBH mass will be smaller than the total mass
derived above since the cluster is made up out of the visible stars,
unseen lower mass stars, possible stellar remnants and the potential
IMBH.  With 300\,\msun\, (seen) but possibly up to $\sim
1000$\,\msun\, of luminous material the mass for the IMBH is then
reduced to 2000 -- 5000\,\msun.  This is much more than the observed
mass of the association, providing a lower limit on its dark mass
component.

\subsection{Simulating IRS\,13E}

With a present density of $\sim4 \times 10^8$\,{\msun}{\rm pc}$^{-3}$,
a collision runaway in IRS\,13E is inevitable, regardless of the
nature of the dark material in the cluster
\citep{1999A&A...348..117P,2004ApJ...604..632G,
2005astro.ph..3129F,2005astro.ph..3130F}. Therefore, even if the
cluster currently does not contain an IMBH, a collision runaway cannot
be prevented if the stars are bound.  We have tested this using N-body
simulations of small clusters of 256 and 1024 stars, with masses drawn
from a Salpeter mass function between 1 and 100\,\msun. These
clusters, with $W_0=6$--9 King model initial density profiles, exactly
filled their Jacobi surfaces, and moved in circular orbits at 0.18\,pc
from the Galactic center. We continued the calculations until the
clusters dissolved.  These simulations lost mass linearly in time,
with a half-mass lifetime of a few 10,000 years, irrespective of the
initial density profile.  This is consistent with the results of
independent symplectic N-body simulations \citep{2005astro.ph..2143L}.
In each of these simulations a minor runaway merger occurred among
roughly a dozen stars, creating runaways of $\aplt 250$\,\msun.  In
another set of larger simulations with 1024--16386 stars, the runaway
mergers were more extreme, with collision rates exceeding one per
century!

We draw two conclusions from these simulations.  If the unseen
material in IRS\,13E consists of normal stars, then (i) the cluster
cannot survive for more than a few $\times10^4$ years, and (ii)
runaway merging is overwhelmingly likely.  If IRS\,13E is bound, a
cluster of normal stars cannot be hidden within it, and the dark
material must ultimately take the form of an IMBH of about
2000--5000\,\msun\, (see also
\cite{2004A&A...423..155M,2005ApJ...625L.111S}).

Thus we argue that the properties of the dark-matter problem in
IRS\,13E could be solved by the presence of a single IMBH of mass
$\sim1000$--5000\,\msun, consistent earlier discussions
\citep{2004A&A...423..155M,2005ApJ...625L.111S}.  The seven observed
stars may in that case be the remnant of a larger star cluster which
has undergone runaway merging, forming the IMBH during core collapse
while sinking toward the Galactic center
\citep{2001ApJ...562L..19E,2003ApJ...593..352P,2005ApJ...628..236G}. According
to this scenario the stars we see are the survivors which have avoided
collision and remained in tight orbits around the IMBH.

Extensive position determinations with the National Radio Astronomy
Observatory's Very Long Baseline Array (VLBA) of Sgr A* over an $\sim
8$ year baseline has revealed that the SMBH in the Galactic center
(assuming 4 million Msun and a distance of 8.0kpc) is about
$7.6\pm0.7$\,km\,s$^{-1}$ \citep{2004ApJ...616..872R}.  An IMBH of
2000--5000\,\msun\, orbiting at a distance of $\sim0.18$\,pc would
create the linear velocity of about 0.15--0.39\,km\,s$^{-1}$ for Sgr
A*, since the orbital velocity of IMBH is $\sim 310$\,km/s and its
mass is $\aplt 1/800$ of the central BH, assuming a circular orbit.
If observations with the VLBA continue with the same accuracy for the
next decade, the IMBH in IRS\,13E can be detected by measuring the
motion of Sgr A*.

\subsection{X-ray and Radio observations of the Galactic center}

X-ray observations may offer a better chance of observing an
individual IMBH near the Galactic center than the VLBA radio
observations discussed in the previous section.  Among the $\sim 2000$
X-ray point sources within 23\,pc of the Galactic center
\citep{2003ApJ...589..225M}, the source CXOGC~J174540.0-290031
\citep{2004astro.ph.12492M}, with $L_{2-8{\rm keV}} \simeq 8.5\times
10^{34}$ erg/s at a projected galactocentric distance of 0.11\,pc, is
of particular interest.  The peak radio intensity of this source is
0.1\,Jansky at 1\,GHz \citep{2005astro.ph..7221B}, which corresponds
to $L_r \sim 8\times 10^{30}$ erg/s at the distance of the Galactic
center. Using the recently proposed empirical relation between X-ray
luminosity, radio flux, and the mass of the accreting black hole
\citep{2003MNRAS.345.1057M},
\begin{equation}
	\log L_r = 7.3 + 0.6 \log L_X + 0.8 \log M_{\rm bh},
\label{Eq:Merloni}\end{equation}
we derive a black hole mass of about 2000\,\msun.

Interestingly, this source has an 7.8 hour
periodicity \citep{2004astro.ph.12492M}, which, if it reflects the
orbital period, would indicate a semi-major axis of $\sim 25$\,\rsun.
The companion to the IMBH would then have a Roche radius of $\sim
1$\,\rsun, consistent with a 1\,{\msun} main-sequence star. Mass
transfer in such a binary would be driven mainly by the emission of
gravitational waves at a rate of $\sim
0.01$\,\msun/Myr \citep{2004MNRAS.355..413P}, which is sufficient to
power an X-ray transient with the observed X-ray luminosity and a duty
cycle on the order of a few percent \citep{2004MNRAS.355..413P}.



\acknowledgements It is a pleasure to thank drs. Clovis Hopman,
Tom Maccarone and Mike Muno for interesting discussions, and
Prof. Ninomiya for the kind hospitality at the Yukawa Institute at
Kyoto University, through the Grants-in-Aid for Scientific Research on
Priority Areas, number 763, ``Dynamics of Strings and Fields,'' from
the Ministry of Education, Culture, Sports, Science and Technology,
Japan.  This work was made possible by financial support from the NASA
Astrophysics Theory Program under grant NNG04GL50G, the Netherlands
Organization for Scientific Research (NWO) under grant 630.000.001,
The Royal Netherlands Academy of Arts and Sciences (KNAW), and the
Netherlands Advanced School for Astronomy (NOVA). Part of the
calculations in this paper were performed on the GRAPE-6 system in
Tokyo and the MoDeStA platform in Amsterdam.


\clearpage

\begin{thebibliography}{}

\bibitem[\protect\astroncite{{Aarseth}}{2003}]{Aarseth2003}
{Aarseth}, S.~A. 2003,
\newblock {Gravitational N-body simulations},
\newblock Cambridge University press, 2003

\bibitem[\protect\astroncite{{Aerts} et~al.}{2004}]{2004A&A...418..639A}
{Aerts}, C., {Lamers}, H.~J.~G.~L.~M., {Molenberghs}, G. 2004, \aap, 418, 639

\bibitem[\protect\astroncite{{Alexander} \&
  {Livio}}{2004}]{2004ApJ...606L..21A}
{Alexander}, T., {Livio}, M. 2004, \apjl, 606, L21

\bibitem[\protect\astroncite{{Baumgardt} \&
  {Makino}}{2003}]{2003MNRAS.340..227B}
{Baumgardt}, H., {Makino}, J. 2003, \mnras, 340, 227

\bibitem[\protect\astroncite{{Baumgardt} et~al.}{2004}]{2004ApJ...613.1143B}
{Baumgardt}, H., {Makino}, J., {Ebisuzaki}, T. 2004, \apj, 613, 1143

\bibitem[\protect\astroncite{{Belkus} et~al.}{2004}]{2004astro.ph.10510B}
{Belkus}, H., {Van Bever}, J., {Vanbeveren}, D. 2004, ArXiv Astrophysics
  e-prints

\bibitem[\protect\astroncite{{Binney} \&
  {Tremaine}}{1987}]{1987gady.book.....B}
{Binney}, J., {Tremaine}, S. 1987,
\newblock Galactic dynamics,
\newblock Princeton, NJ, Princeton University Press, 1987, 747 p.

\bibitem[\protect\astroncite{{Borissova} et~al.}{2005}]{2005A&A...435...95B}
{Borissova}, J., {Ivanov}, V.~D., {Minniti}, D., {Geisler}, D., {Stephens},
  A.~W. 2005, \aap, 435, 95

\bibitem[\protect\astroncite{{Bower} et~al.}{2005}]{2005astro.ph..7221B}
{Bower}, G.~C., {Roberts}, D.~A., {Yusef-Zadeh}, F., {Backer}, D.~C., {Cotton},
  W.~D., {Goss}, W.~M., {Lang}, C.~C., {Lithwick}, Y. 2005, ArXiv Astrophysics
  e-prints

\bibitem[\protect\astroncite{{Brown} et~al.}{2005}]{2005ApJ...622L..33B}
{Brown}, W.~R., {Geller}, M.~J., {Kenyon}, S.~J., {Kurtz}, M.~J. 2005, \apjl,
  622, L33

\bibitem[\protect\astroncite{{Ebisuzaki} et~al.}{2001}]{2001ApJ...562L..19E}
{Ebisuzaki}, T., {Makino}, J., {Tsuru}, T.~G., {Funato}, Y., {Portegies Zwart},
  S., {Hut}, P., {McMillan}, S., {Matsushita}, S., {Matsumoto}, H., {Kawabe},
  R. 2001, \apjl, 562, L19

\bibitem[\protect\astroncite{{Eckart} \& {Genzel}}{1997}]{1997MNRAS.284..576E}
{Eckart}, A., {Genzel}, R. 1997, \mnras, 284, 576

\bibitem[\protect\astroncite{{Figer}}{2004}]{2004astro.ph..9415F}
{Figer}, D.~F. 2004, ArXiv Astrophysics e-prints

\bibitem[\protect\astroncite{{Figer}}{2005}]{2005Natur.434..192F}
{Figer}, D.~F. 2005, \nat, 434, 192

\bibitem[\protect\astroncite{{Figer} \& {Kim}}{2002}]{2002ASPC..263..287F}
{Figer}, D.~F., {Kim}, S.~S. 2002,
\newblock in ASP Conf. Ser. 263: Stellar Collisions, Mergers and their
  Consequences, p.~287

\bibitem[\protect\astroncite{{Figer} et~al.}{1999a}]{1999ApJ...525..750F}
{Figer}, D.~F., {Kim}, S.~S., {Morris}, M., {Serabyn}, E., {Rich}, R.~M.,
  {McLean}, I.~S. 1999a, \apj, 525, 750

\bibitem[\protect\astroncite{{Figer} et~al.}{1999b}]{1999ApJ...514..202F}
{Figer}, D.~F., {McLean}, I.~S., {Morris}, M. 1999b, \apj, 514, 202

\bibitem[\protect\astroncite{{Figer} et~al.}{2002}]{2002ApJ...581..258F}
{Figer}, D.~F., {Najarro}, F., {Gilmore}, D., {Morris}, M., {Kim}, S.~S.,
  {Serabyn}, E., {McLean}, I.~S., {Gilbert}, A.~M., {Graham}, J.~R., {Larkin},
  J.~E., {Levenson}, N.~A., {Teplitz}, H.~I. 2002, \apj, 581, 258

\bibitem[\protect\astroncite{{Freitag} et~al.}{2005a}]{2005astro.ph..3130F}
{Freitag}, M., {Atakan G{\" u}rkan}, M., {Rasio}, F.~A. 2005a, ArXiv
  Astrophysics e-prints

\bibitem[\protect\astroncite{{Freitag} \& {Benz}}{2005}]{2005MNRAS.358.1133F}
{Freitag}, M., {Benz}, W. 2005, \mnras, 358, 1133

\bibitem[\protect\astroncite{{Freitag} et~al.}{2005b}]{2005astro.ph..3129F}
{Freitag}, M., {Rasio}, F.~A., {Baumgardt}, H. 2005b, ArXiv Astrophysics
  e-prints

\bibitem[\protect\astroncite{{G{\" u}ltekin}
  et~al.}{2004}]{2004ApJ...616..221G}
{G{\" u}ltekin}, K., {Miller}, M.~C., {Hamilton}, D.~P. 2004, \apj, 616, 221

\bibitem[\protect\astroncite{{G{\" u}rkan} et~al.}{2004}]{2004ApJ...604..632G}
{G{\" u}rkan}, M.~A., {Freitag}, M., {Rasio}, F.~A. 2004, \apj, 604, 632

\bibitem[\protect\astroncite{{Genzel} et~al.}{2003}]{2003ApJ...594..812G}
{Genzel}, R., {Sch{\" o}del}, R., {Ott}, T., {Eisenhauer}, F., {Hofmann}, R.,
  {Lehnert}, M., {Eckart}, A., {Alexander}, T., {Sternberg}, A., {Lenzen}, R.,
  {Cl{\' e}net}, Y., {Lacombe}, F., {Rouan}, D., {Renzini}, A.,
  {Tacconi-Garman}, L.~E. 2003, \apj, 594, 812

\bibitem[\protect\astroncite{{Gerhard}}{2001}]{2001ApJ...546L..39G}
{Gerhard}, O. 2001, \apjl, 546, L39

\bibitem[\protect\astroncite{{Ghez} et~al.}{2003}]{2003ApJ...586L.127G}
{Ghez}, A.~M., {Duch{\^ e}ne}, G., {Matthews}, K., {Hornstein}, S.~D.,
  {Tanner}, A., {Larkin}, J., {Morris}, M., {Becklin}, E.~E., {Salim}, S.,
  {Kremenek}, T., {Thompson}, D., {Soifer}, B.~T., {Neugebauer}, G., {McLean},
  I. 2003, \apjl, 586, L127

\bibitem[\protect\astroncite{{Ghez} et~al.}{1998}]{1998ApJ...509..678G}
{Ghez}, A.~M., {Klein}, B.~L., {Morris}, M., {Becklin}, E.~E. 1998, \apj, 509,
  678

\bibitem[\protect\astroncite{{Ghez} et~al.}{2000}]{2000Natur.407..349G}
{Ghez}, A.~M., {Morris}, M., {Becklin}, E.~E., {Tanner}, A., {Kremenek}, T.
  2000, \nat, 407, 349

\bibitem[\protect\astroncite{{Gualandris} et~al.}{2005}]{2005MNRAS.363..223G}
{Gualandris}, A., {Zwart}, S.~P., {Sipior}, M.~S. 2005, \mnras, 363, 223

\bibitem[\protect\astroncite{{G{\"u}rkan} \&
  {Rasio}}{2005}]{2005ApJ...628..236G}
{G{\"u}rkan}, M.~A., {Rasio}, F.~A. 2005, \apj, 628, 236

\bibitem[\protect\astroncite{{Hansen} \& {Milosavljevi{\'
  c}}}{2003}]{2003ApJ...593L..77H}
{Hansen}, B.~M.~S., {Milosavljevi{\' c}}, M. 2003, \apjl, 593, L77

\bibitem[\protect\astroncite{{Heavens} et~al.}{2004}]{2004Natur.428..625H}
{Heavens}, A., {Panter}, B., {Jimenez}, R., {Dunlop}, J. 2004, \nat, 428, 625

\bibitem[\protect\astroncite{{Heger} et~al.}{2003}]{2003ApJ...591..288H}
{Heger}, A., {Fryer}, C.~L., {Woosley}, S.~E., {Langer}, N., {Hartmann}, D.~H.
  2003, \apj, 591, 288

\bibitem[Hills(1988)]{1988Natur.331..687H} Hills, J.~G.\ 1988, \nat, 331, 
687 

\bibitem[\protect\astroncite{{Kauffmann} et~al.}{2004}]{2004MNRAS.353..713K}
{Kauffmann}, G., {White}, S.~D.~M., {Heckman}, T.~M., {M{\' e}nard}, B.,
  {Brinchmann}, J., {Charlot}, S., {Tremonti}, C., {Brinkmann}, J. 2004,
  \mnras, 353, 713

\bibitem[\protect\astroncite{{Kim} et~al.}{2000}]{2000ApJ...545..301K}
{Kim}, S.~S., {Figer}, D.~F., {Lee}, H.~M., {Morris}, M. 2000, \apj, 545, 301

\bibitem[\protect\astroncite{{Kim} et~al.}{2004}]{2004ApJ...607L.123K}
{Kim}, S.~S., {Figer}, D.~F., {Morris}, M. 2004, \apjl, 607, L123

\bibitem[\protect\astroncite{{Kudritzki}}{2002}]{2002ApJ...577..389K}
{Kudritzki}, R.~P. 2002, \apj, 577, 389

\bibitem[\protect\astroncite{{Lada} \& {Lada}}{2003}]{2003ARA&A..41...57L}
{Lada}, C.~J., {Lada}, E.~A. 2003, \araa, 41, 57

\bibitem[\protect\astroncite{{Langer} et~al.}{1994}]{1994A&A...290..819L}
{Langer}, N., {Hamann}, W.-R., {Lennon}, M., {Najarro}, F., {Pauldrach},
  A.~W.~A., {Puls}, J. 1994, \aap, 290, 819

\bibitem[\protect\astroncite{{Levin} et~al.}{2005}]{2005astro.ph..2143L}
{Levin}, Y., {Wu}, A.~S.~P., {Thommes}, E.~W. 2005, ArXiv Astrophysics e-prints

\bibitem[\protect\astroncite{{Lombardi} et~al.}{2003}]{2003MNRAS.345..762L}
{Lombardi}, J.~C., {Thrall}, A.~P., {Deneva}, J.~S., {Fleming}, S.~W.,
  {Grabowski}, P.~E. 2003, \mnras, 345, 762

\bibitem[\protect\astroncite{{Lu} et~al.}{2005}]{2005astro.ph..4276L}
{Lu}, J.~R., {Ghez}, A.~M., {Hornstein}, S.~D., {Morris}, M., {Becklin}, E.~E.
  2005, ArXiv Astrophysics e-prints

\bibitem[\protect\astroncite{{Mackey} \& {Gilmore}}{2003}]{2003MNRAS.338...85M}
{Mackey}, A.~D., {Gilmore}, G.~F. 2003, \mnras, 338, 85

\bibitem[\protect\astroncite{{Maillard} et~al.}{2004}]{2004A&A...423..155M}
{Maillard}, J.~P., {Paumard}, T., {Stolovy}, S.~R., {Rigaut}, F. 2004, \aap,
  423, 155

\bibitem[\protect\astroncite{{Makino} et~al.}{2003}]{2003PASJ...55.1163M}
{Makino}, J., {Fukushige}, T., {Koga}, M., {Namura}, K. 2003, \pasj, 55, 1163

\bibitem[\protect\astroncite{{Martins} et~al.}{2004}]{2004A&A...420.1087M}
{Martins}, F., {Schaerer}, D., {Hillier}, D.~J., {Heydari-Malayeri}, M. 2004,
  \aap, 420, 1087

\bibitem[\protect\astroncite{{McMillan} \& {Portegies
  Zwart}}{2004}]{2004astro.ph.12622M}
{McMillan}, S., {Portegies Zwart}, S. 2004, ArXiv Astrophysics
e-prints astro-ph/0412622

\bibitem[\protect\astroncite{{McMillan} \& {Portegies
  Zwart}}{2003}]{2003ApJ...596..314M}
{McMillan}, S.~L.~W., {Portegies Zwart}, S.~F. 2003, \apj, 596, 314

\bibitem[\protect\astroncite{{Menou} et~al.}{2001}]{2001ApJ...558..535M}
{Menou}, K., {Haiman}, Z., {Narayanan}, V.~K. 2001, \apj, 558, 535

\bibitem[\protect\astroncite{{Merloni} et~al.}{2003}]{2003MNRAS.345.1057M}
{Merloni}, A., {Heinz}, S., {di Matteo}, T. 2003, \mnras, 345, 1057

\bibitem[\protect\astroncite{{Merritt} et~al.}{2004}]{2004ApJ...608L..25M}
{Merritt}, D., {Piatek}, S., {Zwart}, S.~P., {Hemsendorf}, M. 2004, \apjl, 608,
  L25

\bibitem[\protect\astroncite{{Merritt} \& {Poon}}{2004}]{2004ApJ...606..788M}
{Merritt}, D., {Poon}, M.~Y. 2004, \apj, 606, 788

\bibitem[\protect\astroncite{{Merritt} \& {Wang}}{2005}]{2005ApJ...621L.101M}
{Merritt}, D., {Wang}, J. 2005, \apjl, 621, L101

\bibitem[\protect\astroncite{{Meynet} \& {Maeder}}{2003}]{2003A&A...404..975M}
{Meynet}, G., {Maeder}, A. 2003, \aap, 404, 975

\bibitem[\protect\astroncite{{Meynet} \& {Maeder}}{2005}]{2005A&A...429..581M}
{Meynet}, G., {Maeder}, A. 2005, \aap, 429, 581

\bibitem[\protect\astroncite{{Morris}}{1993}]{1993ApJ...408..496M}
{Morris}, M. 1993, \apj, 408, 496

\bibitem[\protect\astroncite{{Muno} et~al.}{2003}]{2003ApJ...589..225M}
{Muno}, M.~P., {Baganoff}, F.~K., {Bautz}, M.~W., {Brandt}, W.~N., {Broos},
  P.~S., {Feigelson}, E.~D., {Garmire}, G.~P., {Morris}, M.~R., {Ricker},
  G.~R., {Townsley}, L.~K. 2003, \apj, 589, 225

\bibitem[\protect\astroncite{{Muno} et~al.}{2004}]{2004astro.ph.12492M}
{Muno}, M.~P., {Pfahl}, E., {Baganoff}, F.~K., {Brandt}, W.~N., {Ghez}, A.,
  {Lu}, J., {Morris}, M.~R. 2004, ArXiv Astrophysics e-prints

\bibitem[\protect\astroncite{{Najarro} et~al.}{1997}]{1997A&A...325..700N}
{Najarro}, F., {Krabbe}, A., {Genzel}, R., {Lutz}, D., {Kudritzki}, R.~P.,
  {Hillier}, D.~J. 1997, \aap, 325, 700

\bibitem[\protect\astroncite{{Nayakshin} \&
  {Cuadra}}{2004}]{2004astro.ph..9541N}
{Nayakshin}, S., {Cuadra}, J. 2004, ArXiv Astrophysics e-prints

\bibitem[\protect\astroncite{{Nayakshin} \&
  {Sunyaev}}{2005}]{2005astro.ph..7687N}
{Nayakshin}, S., {Sunyaev}, R. 2005, ArXiv Astrophysics e-prints

\bibitem[\protect\astroncite{{Paumard} et~al.}{2001}]{2001A&A...366..466P}
{Paumard}, T., {Maillard}, J.~P., {Morris}, M., {Rigaut}, F. 2001, \aap, 366,
  466

\bibitem[\protect\astroncite{{Portegies Zwart}
  et~al.}{2004a}]{2004Natur.428..724P}
{Portegies Zwart}, S.~F., {Baumgardt}, H., {Hut}, P., {Makino}, J., {McMillan},
  S.~L.~W. 2004a, \nat, 428, 724

\bibitem[\protect\astroncite{{Portegies Zwart}
  et~al.}{2004b}]{2004MNRAS.355..413P}
{Portegies Zwart}, S.~F., {Dewi}, J., {Maccarone}, T. 2004b, \mnras, 355, 413

\bibitem[Portegies Zwart et al.(2003)]{2003ApJ...593..352P} Portegies 
Zwart, S.~F., McMillan, S.~L.~W., \& Gerhard, O.\ 2003, \apj, 593, 352 

\bibitem[\protect\astroncite{{Portegies Zwart}
  et~al.}{1999}]{1999A&A...348..117P}
{Portegies Zwart}, S.~F., {Makino}, J., {McMillan}, S.~L.~W., {Hut}, P. 1999,
  \aap, 348, 117

\bibitem[\protect\astroncite{{Portegies Zwart}
  et~al.}{2001a}]{2001ApJ...546L.101P}
{Portegies Zwart}, S.~F., {Makino}, J., {McMillan}, S.~L.~W., {Hut}, P. 2001a,
  \apjl, 546, L101

\bibitem[\protect\astroncite{{Portegies Zwart} \&
  {McMillan}}{2002}]{2002ApJ...576..899P}
{Portegies Zwart}, S.~F., {McMillan}, S.~L.~W. 2002, \apj, 576, 899

\bibitem[\protect\astroncite{{Portegies Zwart}
  et~al.}{2001b}]{2001MNRAS.321..199P}
{Portegies Zwart}, S.~F., {McMillan}, S.~L.~W., {Hut}, P., {Makino}, J. 2001b,
  \mnras, 321, 199

\bibitem[\protect\astroncite{{Quinlan} \&
  {Shapiro}}{1990}]{1990ApJ...356..483Q}
{Quinlan}, G.~D., {Shapiro}, S.~L. 1990, \apj, 356, 483

\bibitem[\protect\astroncite{{Reid} \&
  {Brunthaler}}{2004}]{2004ApJ...616..872R}
{Reid}, M.~J., {Brunthaler}, A. 2004, \apj, 616, 872

\bibitem[\protect\astroncite{{Sandqvist} et~al.}{2003}]{2003A&A...402L..63S}
{Sandqvist}, A., {Bergman}, P., {Black}, J.~H., {Booth}, R., {Buat}, V.,
  {Curry}, C.~L., {Encrenaz}, P., {Falgarone}, E., {Feldman}, P., {Fich}, M.,
  {Floren}, H.~G., {Frisk}, U., {Gerin}, M., {Gregersen}, E.~M., {Harju}, J.,
  {Hasegawa}, T., {Hjalmarson}, {\AA}., {Johansson}, L.~E.~B., {Kwok}, S.,
  {Larsson}, B., {Lecacheux}, A., {Liljestr{\" o}m}, T., {Lindqvist}, M.,
  {Liseau}, R., {Mattila}, K., {Mitchell}, G.~F., {Nordh}, L., {Olberg}, M.,
  {Olofsson}, A.~O.~H., {Olofsson}, G., {Pagani}, L., {Plume}, R.,
  {Ristorcelli}, I., {Sch{\' e}ele}, F.~v., {Serra}, G., {Tothill}, N.~F.~H.,
  {Volk}, K., {Wilson}, C.~D., {Winnberg}, A. 2003, \aap, 402, L63

\bibitem[\protect\astroncite{{Sch{\" o}del} et~al.}{2005}]{2005ApJ...625L.111S}
{Sch{\" o}del}, R., {Eckart}, A., {Iserlohe}, C., {Genzel}, R., {Ott}, T. 2005,
  \apjl, 625, L111

\bibitem[\protect\astroncite{{Spinnato} et~al.}{2003}]{2003MNRAS.344...22S}
{Spinnato}, P.~F., {Fellhauer}, M., {Portegies Zwart}, S.~F. 2003, \mnras, 344,
  22

\bibitem[\protect\astroncite{{Spitzer} \& {Hart}}{1971}]{1971ApJ...166..483S}
{Spitzer}, L.~J., {Hart}, M.~H. 1971, \apj, 166, 483

\bibitem[\protect\astroncite{{Stolte} et~al.}{2005}]{2005ApJ...628L.113S}
{Stolte}, A., {Brandner}, W., {Grebel}, E.~K., {Lenzen}, R., {Lagrange}, A.-M.
  2005, \apjl, 628, L113

\bibitem[\protect\astroncite{{Tamblyn} \& {Rieke}}{1993}]{1993ApJ...414..573T}
{Tamblyn}, P., {Rieke}, G.~H. 1993, \apj, 414, 573

\bibitem[\protect\astroncite{{van Boekel} et~al.}{2003}]{2003A&A...410L..37V}
{van Boekel}, R., {Kervella}, P., {Sch{\" o}ller}, M., {Herbst}, T.,
  {Brandner}, W., {de Koter}, A., {Waters}, L.~B.~F.~M., {Hillier}, D.~J.,
  {Paresce}, F., {Lenzen}, R., {Lagrange}, A.-M. 2003, \aap, 410, L37

\bibitem[\protect\astroncite{{Vink} et~al.}{2000}]{2000A&A...362..295V}
{Vink}, J.~S., {de Koter}, A., {Lamers}, H.~J.~G.~L.~M. 2000, \aap, 362, 295

\bibitem[\protect\astroncite{{Vink} et~al.}{2001}]{2001A&A...369..574V}
{Vink}, J.~S., {de Koter}, A., {Lamers}, H.~J.~G.~L.~M. 2001, \aap, 369, 574

\bibitem[\protect\astroncite{{von Zeipel}}{1924}]{1924MNRAS..84..665V}
{von Zeipel}, H. 1924, \mnras, 84, 665

\bibitem[\protect\astroncite{{Yu} \& {Tremaine}}{2003}]{2003ApJ...599.1129Y}
{Yu}, Q., {Tremaine}, S. 2003, \apj, 599, 1129

\bibitem[\protect\astroncite{{Zhang} \& {Fall}}{1999}]{1999ApJ...527L..81Z}
{Zhang}, Q., {Fall}, S.~M. 1999, \apjl, 527, L81

\end{thebibliography}
\end{document}